\documentclass[useAMS,usenatbib]{mn2e}
\usepackage{psfig}
\usepackage[a4paper]{hyperref}
\usepackage{amssymb}
\usepackage{aas_macros}

\newcommand{\rmsub}[2]{#1_{\rm #2}}

\begin{document}

\title[The transiting planet WASP-10b]{WASP-10b: a 3M$_J$, gas-giant planet transiting a late-type K star}
\author[D.J. Christian et al.]
{
D.J. Christian$^{1,17}$\thanks{E-mail: d.christian@qub.ac.uk},
N.P. Gibson$^{1}$,
E.K. Simpson$^{1}$,
R.A. Street$^{1,2}$,
I. Skillen$^{3}$,
\newauthor
D. Pollacco$^{1}$,
A. Collier Cameron$^{4}$,
Y.C. Joshi$^{1}$,
F.P. Keenan$^{1}$,
H.C. Stempels$^{4}$,
\newauthor
C.A. Haswell$^{5}$,
K. Horne$^{4}$,
D.R. Anderson$^{5}$, 
S. Bentley$^{6}$,  
F. Bouchy$^{7,8}$, 
W.I. Clarkson$^{5,16}$,
\newauthor
B. Enoch$^{5}$,   
L. Hebb$^{4}$, 
G. H\'ebrard$^{7}$, 
C. Hellier$^{6}$,  
J. Irwin$^{9}$,  
S.R. Kane$^{10}$,
\newauthor
T.A. Lister$^{2,4,6}$,
B. Loeillet$^{11}$, 
P. Maxted$^{6}$,
M. Mayor$^{12}$, 
I. McDonald$^{6}$, 
C. Moutou$^{11}$, 
\newauthor
A.J. Norton$^{5}$,  
N. Parley$^{5}$,  
F. Pont$^{12,13}$, 
D. Queloz$^{12}$, 
R. Ryans$^{1}$,
B. Smalley$^{6}$, 
\newauthor
A.M.S. Smith$^{4}$,
I. Todd$^{1}$,  
S. Udry$^{12}$, 
R.G. West$^{14}$,
P.J. Wheatley$^{15}$,
D.M. Wilson$^{6}$ 
\\
\\
$^{1}$Astrophysics Research Centre, School of Mathematics \&\ Physics, Queen's University, University Road, Belfast, BT7 1NN, UK\\
$^{2}$Las Cumbres Observatory, 6740 Cortona Dr. Suite 102, Santa Barbara, CA 93117, USA\\
$^{3}$Isaac Newton Group of Telescopes, Apartado de Correos 321, E-38700 Santa Cruz de la Palma, Tenerife, Spain \\
$^{4}$School of Physics and Astronomy, University of St Andrews, North Haugh, St Andrews, Fife KY16 9SS, UK\\
$^{5}$Department of Physics and Astronomy, The Open University, Milton Keynes, MK7 6AA, UK\\
$^{6}$Astrophysics Group, Keele University, Staffordshire, ST5 5BG\\
$^{7}$Institut d'Astrophysique de Paris, CNRS (UMR 7095) --  Universit\'e Pierre \&\ Marie Curie, 98$^{bis}$ bvd. Arago, 75014 Paris, France\\
$^{8}$Observatoire de Haute-Provence, 04870 St Michel l'Observatoire, France\\
$^{9}$ Harvard/Smithsonian Center for Astrophysics, 60 Garden Street, Cambridge, MA 02138, USA\\
$^{10}$Michelson Science Center, Caltech, MS 100-22, 770 South Wilson Avenue
Pasadena, CA 91125, USA \\
$^{11}$Laboratoire d'Astrophysique de Marseille, BP 8, 13376 Marseille Cedex 12, France\\
$^{12}$Observatoire de Gen\`eve, Universit\'e de Gen\`eve, 51 Ch. des Maillettes, 1290 Sauverny, Switzerland\\
$^{13}$ School of Physics, University of Exeter, Stocker Road, Exeter EX4 4QL\\
$^{14}$Department of Physics and Astronomy, University of Leicester, Leicester, LE1 7RH, UK\\
$^{15}$Department of Physics, University of Warwick, Coventry CV4 7AL, UK\\
$^{16}$STScI, 3700 San Martin Drive, Baltimore, MD 21218, USA\\
$^{17}$Department of Physics and Astronomy, California State University Northridge, 18111 Nordhoff Street, Northridge, CA 91330-8268, USA\\
}

 \date{Accepted 2008 October ??. Received 2008 September ??; in original form 2008 May ??}

\pagerange{\pageref{firstpage}--\pageref{lastpage}} \pubyear{2008}

\maketitle

\label{firstpage}

\begin{abstract}

We report the discovery of WASP-10b, a new transiting 
extrasolar planet (ESP) discovered by the WASP Consortium and confirmed
using NOT FIES and SOPHIE radial velocity data.  A
3.09 day period, 29 mmag transit depth, and 2.36 hour duration
are derived for WASP-10b using WASP
and high precision photometric observations.
Simultaneous fitting to the photometric and radial velocity data using 
a Markov-chain Monte Carlo procedure leads to a planet radius of 
1.28R$_J$, a mass of 2.96M$_J$ and eccentricity of $\approx$0.06. 
WASP-10b is one of the more massive transiting ESPs, and we compare its
characteristics to the current sample of transiting ESP, where there is 
currently little information for masses greater than $\approx$2M$_J$
and non-zero eccentricities.  
WASP-10's host star, GSC 2752-00114
(USNO-B1.0 1214-0586164)
is among the fainter stars in the WASP sample, with V=12.7 and a spectral
type of K5.  This result shows promise for future late-type dwarf star surveys.

\end{abstract}


\section{Introduction}
Photometric transit observations of extrasolar planets (ESP) are
important because the transit strongly constrains their orbital
inclination and allows accurate physical parameters 
for the planet to be derived. 
Their mass-radius relation allows us to probe their
internal structure and is vital to our understanding of orbital
migration and planetary formation.  
The radial velocity measurements which are used to confirm a candidate
transiting ESP also provide more complete information  on the
orbital eccentricity.

As wide field photometric transit surveys have collected additional
sky and temporal coverage, and understood 
their noise components \citep{cameron2007mcmc, smith07}, 
the number of transiting ESP has grown to over 50
in line with earlier predictions \citep{h1}.
Recently one
such survey, SuperWASP \citep{p1} published its first 5 
confirmed ESP, all of which have periods of less than 3 days
\citep{A08, c4, P08, W08}, and reported an additional 10\footnote{http://www.inscience.ch/transits/}
\citep{Hel08, Hebb08, yog08, West08}.
SuperWASP is performing a "shallow-but-wide"
transit search, designed to find planets that are not only
sufficiently bright ($9 < V < 13$) for high-precision radial velocity follow-up
to be feasible on telescopes of moderate aperture, but also for
detailed studies such as transmission spectroscopy during transits.
Details of the WASP project and observatory infrastructure are 
described in \citet{p1}.

In this paper we present the WASP photometry of 
1SWASP~J231558.30+312746.4 (GSC 2752-00114),
higher precision photometric follow-up observations with the MERCATOR and 
Tenagra telescopes,
and high precision radial velocity observations with the Nordic Optical Telescope new 
FIbre-fed Echelle Spectrograph (FIES) and the OHP SOPHIE collaboration. 
These observations lead to the discovery and confirmation of a new,
relatively high mass, gas-giant exoplanet, WASP-10b. 

\begin{figure}
\psfig{file=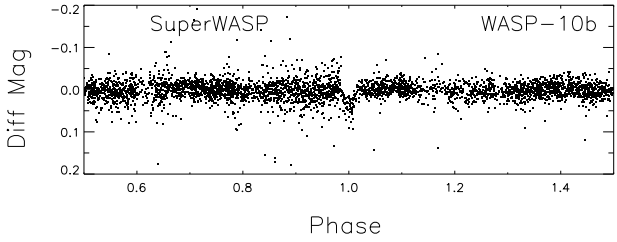,width=8.0cm}
\caption{ 
(a.) The {\it top} panel shows the SuperWASP light curve for WASP-10b (1SWASP~J231558.30+312746.4). 
All the data (apart from that from SuperWASP-N) were averaged 
in 300\,second bins. The data were phased using the ephemeris, 
$T_O = 2454357.85808 ^{+ 0.00041}_{- 0.00036}$
$P = 3.09276$ days. 
}
\end{figure}

\section{Observations}
 
1SWASP~J231558.30+312746.4 (GSC 2752-00114)   
was monitored by SuperWASP-N starting in 2004.
SuperWASP is a multi-camera telescope system with SuperWASP-North 
located in La Palma and consisting of 8  
Canon 200-mm f/1.8 lenses each coupled to e2v 2048 x 2048 pixel
back illuminated CCDs.
This combination of lens and camera yields a field of view of 
7.8$^o$ $\times$ 7.8$^o$ 
with an angular size of 14.2$\arcsec$ per pixel. 
During 2004, SuperWASP was run with 4 or 5 cameras 
as the operations moved from commissioning to routine automated observing.
We show the WASP-10b SuperWASP light curve in Figure 1.

\subsection{ Higher Precision Photometry}

We obtained photometry of WASP-10 with the MEROPE instrument on the 1.2m MERCATOR Telescope in V-band on 1 September 2007.  
Only a partial transit was observed due to uncertainties in the
period and epoch from the SuperWASP data. Observations were in the V-band with 
2$\times$2 binning over $\sim$2.9 hours. Despite clear conditions, exposure times were varied from 25-30s to account for changes in seeing and to keep below
the saturation limits of the chip. This allowed 170 images to be taken. There was a
drift of only ~1 binned pixel in x and y on the chip during the run.
The MEROPE images were first de-biased and flat fielded with combined twilight flats using IRAF and the apphot package to obtain aperture photometry of the target and
5 nearby companion stars using a 10 pixel radius. Finally, the light curve 
was extracted and normalized to reveal a depth of $\sim$33 mmag.

Further observations of WASP-10 were taken 
as part of an observing program sponsored by  
the Las Cumbres Observatory Global Telescope  
Network\footnote{www.lcogt.net} on the Tenagra II, 0.81m F7  
Ritchey-Chretien telescope sited in the Sonora desert in S. Arizona,  
USA.  The science camera contains a 1k $\times$ 1k SITe CCD with a pixel scale  
of 0.87 arcsec/pixel and a field of view of 14.8$\arcmin \times 14.8\arcmin$.  
The  filter set is the standard Johnson/Cousins/Bessel UBVRI set and
the data presented here have been taken in I band.  

Calibration frames were obtained automatically every twilight, and the  
data were de-biased and flat-fielded using the calibration section of  
the SuperWASP pipeline.  Object detection and aperture photometry was then  
performed using daophot \citep{Stet87} within
IRAF.  Differential photometry was derived from a selection of  
typically 5 to 10 comparison stars within the frame.

These confirmed the object had a sharp egress with an amplitude of 0.033$\pm$0.001mag. 
 The MERCATOR V and Tenagra I light curves show consistent transit depths, 
confirming that the companion is a transiting ESP.

\subsection{Spectroscopic Follow-up}
\label{section:radvel}
 We obtained high precision radial velocity (RV) follow-up observations of
WASP-10 with the 2.5m Nordic Optical Telescope (NOT) new FIbre-fed 
Echelle Spectrograph (FIES),
supplemented with observations from the Observatoire de Haute-Provence's 
1.93\,m telescope and the SOPHIE spectrograph \citep{b1}. 
We present a summary of the FIES and SOPHIE RV data in Table~\ref{tab:rvspec}.
 
 \begin{table}
\caption[]{Journal of radial-velocity measurements for WASP-10 (1SWASP~J231558.30+312746.4, USNO-B1.0 1214-0586164). 
Stellar coordinates are for the photometric apertures; 
the USNO-B1.0 number denotes 
the star for which the radial-velocity measurements were secured.  
The quoted uncertainties in the radial velocity errors include 
components due to photon noise (Section~\ref{section:radvel}) and 10 m\,s$^{-1}$ of jitter 
(Section~\ref{section:mcmc}) added in quadrature. 
}
\begin{center}
\begin{tabular}{lcccc}
BJD & $\rmsub{t}{exp}$ &$\rmsub{V}{r}$  \\
     &       (s)        &  km s$^{-1}$    \\
\hline
{\em FIES} \\

2454437.540  & 2400  &   -11.028  $\pm$ 0.026   &   \\
2454463.377  & 2400  &   -11.941  $\pm$ 0.030   &   \\
2454465.342  & 2400  &   -11.003  $\pm$ 0.018   &   \\
2454466.335  & 2400  &   -11.804  $\pm$ 0.021   &   \\
2454490.329  & 2400  &   -11.013  $\pm$ 0.143    &   \\
2454490.358  & 2400  &   -10.990  $\pm$ 0.120    &   \\
2454491.340  & 2400  &   -11.955  $\pm$ 0.024   &   \\


{\em SOPHIE}\\
2454340.569 &	  3300   & -11.657 $\pm$	0.008   \\
2454342.505 &	  3600   & -11.575 $\pm$	0.011   \\
2454508.262 &	 1680    & -11.027 $\pm$	0.014    \\
2454509.268 &	 1680    & -11.336 $\pm$	0.017   \\
2454510.276 &	 1680    & -11.990 $\pm$	0.020    \\
2454511.262 &	 1680    & -11.135 $\pm$	0.016    \\
2454512.262 &	 1680    & -11.244 $\pm$ 	0.016    \\
%

\hline\\
\end{tabular}
\end{center}
\label{tab:rvspec}
\end{table}

\subsubsection{NOT and FIES}

 Spectroscopic observations were obtained using the 
new FIES spectrograph mounted on the NOT Telescope. 
A total of seven radial velocity points were obtained 
during 2 December 2007, 28--31 December 2007 and 24--25 January 2008. 
WASP-10 required observations with an exposure time of 2400s due its 
relative faintness (V=12.7) yielding a peak signal-to-noise ratio per 
resolution element of  $\approx$60--70 in the H$\alpha$ region. FIES was used in 
medium resolution mode with R=46000 with simultaneous ThAr 
calibration. We used the bespoke data reduction package 
FIEStool\footnote{http://www.not.iac.es/instruments/fies/fiestool/FIEStool.html} to 
extract the spectra and a specially developed IDL line-fitting code 
to obtain radial velocities with a precision of 15--25 m s$^{-1}$ (except for the
poor night of 24 January 2008, JD 2454490).

\subsubsection{OHP 1.9\,m and SOPHIE}
 Additional radial velocity measurements were taken for 
WASP-10 on 2007 August 29 and 30, and again between
2008 Feb 11 and 15 with the OHP 1.93\,m telescope and the SOPHIE 
spectrograph \citep{b1}, a total of 7 usable spectra were acquired.
We used SOPHIE in its high efficiency mode, acquiring simultaneous star 
and sky spectra through separate fibers
with a resolution of R=40000. Thorium-Argon calibration images were taken 
at the start and end 
of each night, and at 2- to 3-hourly intervals throughout the night.  
The radial-velocity drift never exceeded 2-3 m\,s$^{-1}$, even on a night-to-night 
basis.  Although errors for each radial velocity measurement are limited by the
photon-noise.  Thus, the average radial velocity error is $\approx$ 14  m\,s$^{-1}$ and
includes the 2-3 m\,s$^{-1}$ systematic error and the contribution from the photon-noise.
Typical signal-to-noise ratio estimates for each spectra were $\approx$30 (near
5500 \AA).
The SOPHIE WASP-10 spectra were cross-correlated 
against a K5V template provided by the SOPHIE control and reduction software.
Typical FWHM and contrasts for these spectra were $\approx$10.2--10.4 km s$^{-1}$ 
and 30--31\%, respectively. The cross-correlation techniques and derivation 
of errors in the radial velocity measurements are presented in \citet{P08}.

\section{Results and Analysis}
\subsection{Stellar parameters}

\begin{table}
\caption[]{Stellar parameters for WASP-10. The last 5 parameters were derived 
from the SME analysis of the FIES spectroscopy.}
\label{tab:star}
\begin{center}
\begin{tabular}{cc}
 Parameter   & WASP-10 \\
 \hline\\
RA (J2000)   & 23 15 58.3 \\
Dec (J2000)  & +31 27 46.4 \\
V  &  12.7  \\
distance & 90$\pm$20 pc \\  
$T_{\rm eff}$  & $4675 \pm 100$ K \\
log\,$g$          & $4.40 \pm 0.20$     \\
$[$M/H$]$      & $0.03 \pm 0.20$  \\ 
$v$\,sin\,$i$    & $<$6 km\,s$^{-1}$          \\
$v_{\rm rad}$   & -11.44 $\pm$ 0.03 km\,s$^{-1}$   \\   
\hline\\
\end{tabular}
\end{center}
\end{table}

We merged all available WASP-10 FIES spectra into one
high-quality spectrum in order to perform a detailed spectroscopic analysis of the stellar atmospheric properties. Radial velocity signatures were carefully removed during
the process. This merged spectrum was then continuum-normalized with a low
order polynomial to retain the shape of the broadest spectral features. The
total signal-to-noise ratio of the combined spectrum was $\approx$180 per pixel. 
We were not able to include the SOPHIE spectra in this analysis, 
because these spectra were obtained in the HE (high-efficiency) mode, which is known to suffer from problems with removal of the blaze function.

As previously undertaken for our analysis of WASP-1 \citep{s2}, and WASP-3 
\citep{P08} we employed the methodology of \citet{v2}, using the
same tools, techniques and model atmosphere grid. We used the package {\it
Spectroscopy Made Easy} ({\sc sme})  \citep{v1},
which combines spectral synthesis with multidimensional $\chi^2$ minimization to determine which atmospheric parameters best reproduce the observed spectrum of WASP-10 (effective temperature $T_{\rm eff}$, surface gravity $\log g$, metallicity
[M/H], projected radial velocity $v \sin i$, systemic radial velocity $v_{\rm
rad}$, microturbulence $v_{\rm mic}$ and the macroturbulence $v_{\rm mac}$). 

The four spectral regions we used in our analysis are: (1) 5160--5190 {\AA}, covering the gravity-sensitive
Mg b triplet (2) 5850--5950 {\AA}, with the temperature and gravity-sensitive Na {\i} D doublet; (3) 6000-6210 {\AA}, containing a wealth of different metal lines, providing leverage on the metallicity, and (4)
6520--6600 {\AA}, covering the strongly temperature-sensitive H-alpha line. In addition we analyzed a small
region around the Li {\sc i} 6708 line to possibly derive a lithium abundance, but no Li {\sc i} 6708 was
detected for WASP-10. The parameters we obtained from this analysis are listed in
Table~\ref{tab:star}.
In addition to the spectral analysis, we also use available photometry 
(from NOMAD, TASS4 and CMC14 catalogues), plus 2MASS to 
estimate the effective temperature using the Infrared Flux Method 
\citep{bs77}. This yields T$_{\rm eff}$ = 4650 $\pm$ 120 K, which is 
in agreement with the spectroscopic analysis and a spectral type of K5.
The characteristics of WASP-10 are also given in Table~\ref{tab:star}.

\begin{figure}[t]
\begin{center}
\psfig{figure=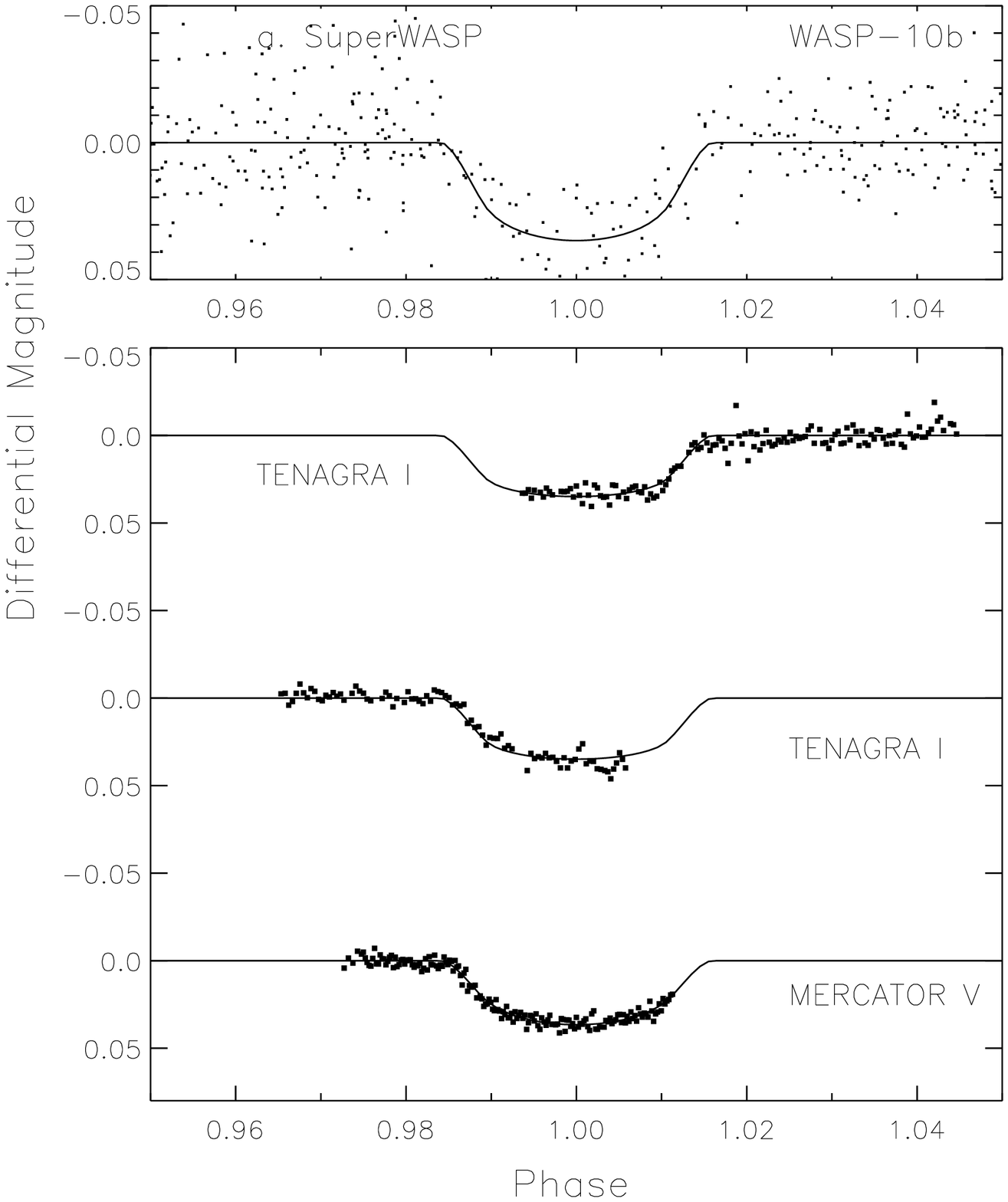,width=8.2cm,angle=0}
\medskip
\psfig{figure=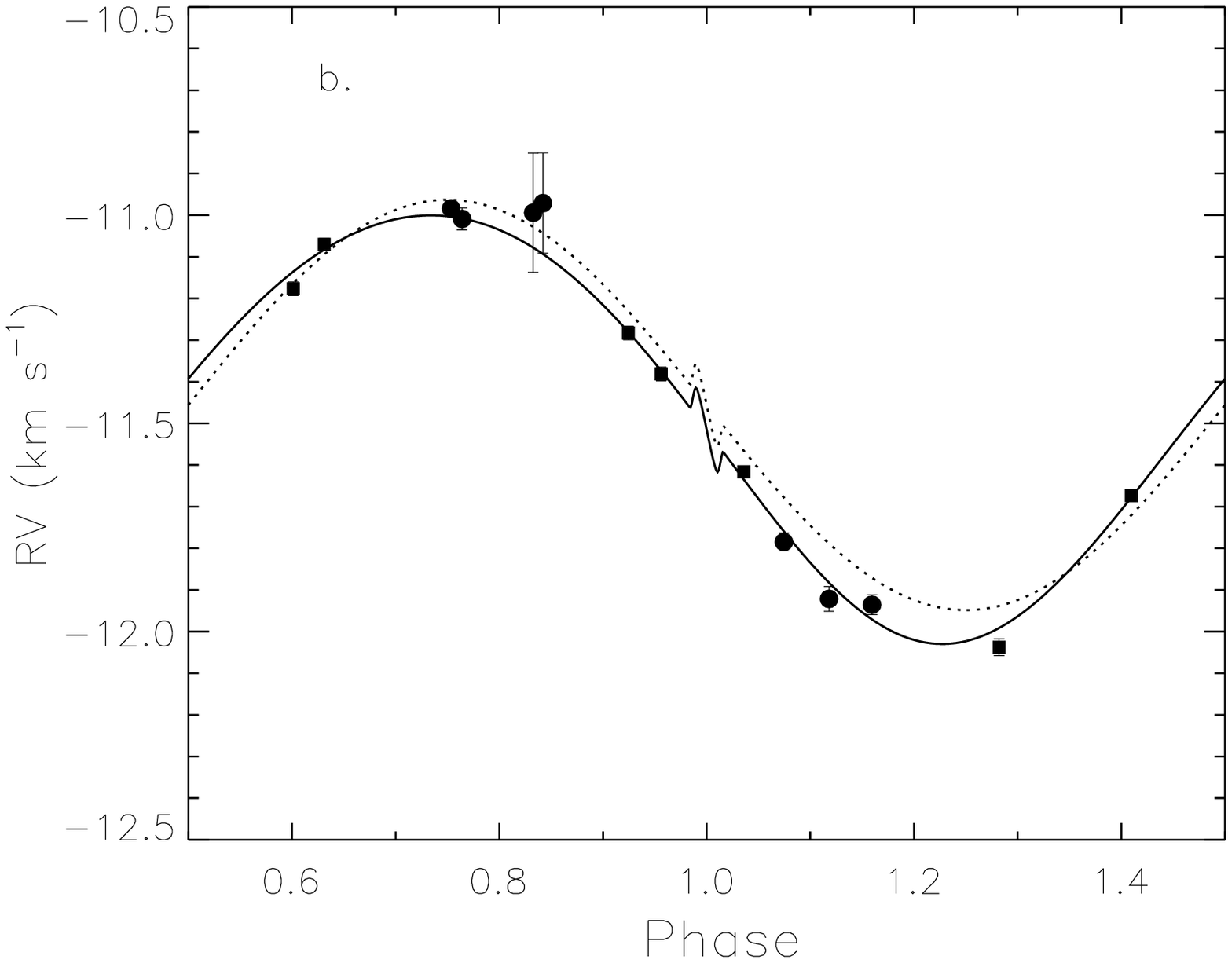,width=8.2cm}  
\caption[]{
Simultaneous Markov-chain Monte Carlo (MCMC) solutions to the WASP-10 
photometry and radial velocity data. 
a. {\it Top} panels show the MCMC solutions to the combined SuperWASP-N, MERCATOR $V$, and Tenagra $I$ band photometry. 
b. The {\it lower} panel
shows the MCMC solution to the FIES + SOPHIE radial velocity (RV) data. 
(FIES RVs are shown as filled circles and SOPHIE as filled squares).
The model fit to the RV data includes orbital eccentricity (solid line), and
for a circular orbit (dashed line). 
Both  RV models also include the Rossiter-McLaughlin effect, which is
small for this system given the low $v sin i$ of the host star ($<$6\,km\,s$^{-1}$).
}
\label{fig:sol_rvlc}
\end{center}
\end{figure}

\subsection{Markov-chain Monte Carlo analysis}
\label{section:mcmc}

Transit timing and the radial-velocity measurements 
provide detailed information about the orbit.
We modelled WASP-10b's transit photometry and the reflex
motion of the host star simultaneously  
using the Markov-chain Monte-Carlo algorithm
described in detail by \citet{cameron2007mcmc},
and the same techniques that
were applied to WASP-3 by \citet{P08}  to which we refer the reader
for more details.

We find  WASP-10b to have a radius 
$1.28 ^{+ 0.08 }_{- 0.09 } R_J$, mass of $2.96 ^{+ 0.22 }_{- 0.17 } M_J$
and a significant non-zero eccentricity of $0.059 ^{ + 0.014}_{ -0.004}$ .
The best fit solution for the MCMC model for a circular orbit
(e = 0) has a $\chi^2$ 55 higher than the solution with non-zero eccentricity,
and thus, the eccentricity is significant at $>$ 99.6\% confidence level using the $F$ test.
The values of the parameters of the optimal solution are given, 
together with their associated 1-$\sigma$ confidence intervals, 
in Table~\ref{tab:params}.
The FIES+SOPHIE radial-velocity data measurements are plotted in Figure~\ref{fig:sol_rvlc}
together with the best-fitting global fit to the SuperWASP-N, MERCATOR, and Tenagra 
transit photometry. 

\begin{figure}
\begin{center}
\psfig{figure=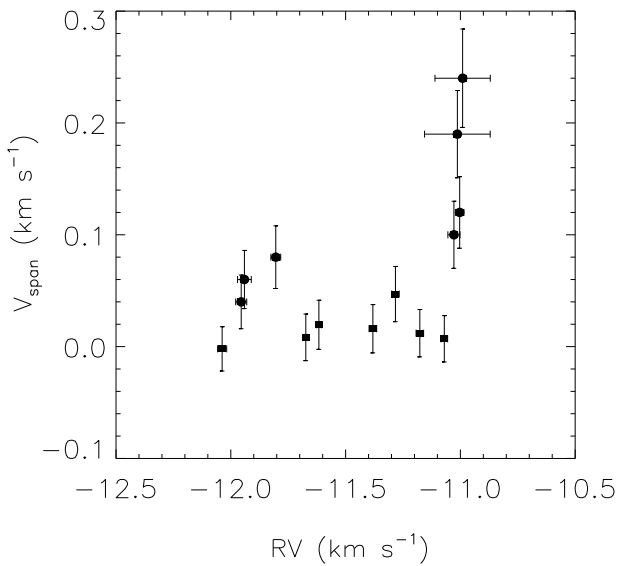,width=8.0cm}
\caption[]{Line bisectors as a function of radial velocity (RV) for WASP-10.
Plot symbols are the same as Figure 2.
Analysis of these line-bisectors for WASP-10 does not show a correlation between
the bisector velocity (V$_{span}$) and stellar radial-velocity (see text).  
}
\label{fig:line_bisector}
\end{center}
\end{figure}

\subsection{Line-bisector variation}
Line bisectors have been shown to be a powerful diagnostic in distinguishing
true extra-solar planets from blended and eclipsing stellar systems chromospheric activity \citep{bisectq1}.
\citet{torr04} showed,  that for OGLE-TR-33 line asymmetries which changed 
with a 1.95 day period, it was a blended system. 
From the cross-correlation function (CCF) we obtained the
line bisectors and these are plotted, as a function of RV, 
in Figure~\ref{fig:line_bisector}.

We quantified the significance of the bisector variation as follows. We
determined the inverse-variance weighted averages of the RV and bisector span as
$$\hat{v} = \frac{\sum_i v_i w_i}{\sum_i w_i};~~~~
\hat{b} = \frac{\sum_i b_i w_i}{\sum_i w_i}$$
where the $v_i$ and $b_i$ are the RV and span bisector values respectively and the weights $w_i$ are
the inverse variances of the individual bisector measurements. The uncertainty in the span bisector
is assumed to be 2.5 times the uncertainty on the RV in our data. If we define $x_i = v_i - \hat{v}$
and $y_i = b_i - \hat{b}$, then the slope is determined as
$$\hat{a} = \frac{\sum_i x_i y_i w_i}{\sum_i x_i^2 w_i};~~~~
{\rm Var}(\hat{a}) = \frac{1}{\sum_i x_i^2 w_i}$$
The value of the scaling factor $\hat{a}$ is determined with signal-to-noise ratio 
$${\rm SNR} = \frac{\hat{a}}{\sqrt {Var(\hat{a})}}$$
We obtain ${\rm SNR}$ = 1.16, indicating a non-significant correlation between
the bisector span and radial velocity variations.
This demonstrates that the
cross-correlation function remains symmetric, and that the radial-velocity 
variations are not likely to be caused by line-of-site binarity or 
stellar activity and indicate WASP-10b is an exoplanet.

\begin{table}
\caption[]{WASP-10 system parameters and 1-$\sigma$ error limits derived
from MCMC analysis.}
\label{tab:params}
\begin{tabular}{lcc}
\hline\\
Parameter & Symbol & Value  \\
\hline\\
Transit epoch (BJD) & $ T_0  $ & $ 2454357.85808 ^{+0.00041}_{-0.00036}$  days \\

Orbital period & $ P  $ & $ 3.0927636^{+0.0000094}_{-0.000021} $  days \\

Planet/star area ratio  & $ (R_p/R_s)^2 $ & $ 0.029^{+0.001}_{-0.001} $   \\
 
Transit duration & $ t_T $ & $ 0.098181^{+0.0019 }_{- 0.0015} $  days \\

Impact parameter & $ b $ & $ 0.568^{+0.054}_{-0.084}  $  $R_*$ \\

  &    &        \\
Stellar reflex velocity & $ K_1 $ & $ 0.5201^{+0.0084}_{-0.010} $  km s$^{-1}$ \\
  
Centre-of-mass velocity  & $ \gamma $ & $ -11.4854^{+0.0012}_{-0.0034} $  km s$^{-1}$ \\
  
Orbital semimajor axis & $ a $ & $ 0.0369^{+0.0012}_{-0.0014} $  AU \\
  &    &       \\

Orbital inclination & $ I $ & $ 86.9 ^{+ 0.6 }_{- 0.5 } $  degrees \\

Orb. eccentricity & $\epsilon$ & $0.059 ^{ + 0.014}_{ -0.004}$    \\

Arg. periastron   & $\omega$   &  $2.917 ^{ +  0.222}_{ - 0.245}$   (rad)  \\
&    &       \\

Stellar mass & $ M_* $ & $ 0.703^{ + 0.068}_{-0.080} $  $M_\odot$ \\
Stellar radius & $ R_* $ & $ 0.775 ^{+ 0.043 }_{- 0.040 } $  $R_\odot$ \\
Stellar surface gravity & $ \log g_* $ & $ 4.51 ^{+0.06}_{ - 0.05}$  (CGS) \\

Stellar density & $ \rho_* $ & $ 1.51 ^{+ 0.25 }_{- 0.20 } $  $\rho_\odot$ \\
&    &       \\
Planet radius & $ R_p $ & $ 1.28 ^{+ 0.077 }_{- 0.091 } $  $R_J$ \\
Planet mass & $ M_p $ & $ 2.96 ^{+ 0.22 }_{- 0.17 } $  $M_J$ \\

Planetary surface gravity & $ \log g_p $ & $ 3.62 \pm 0.06 $  (CGS) \\

Planet density & $ \rho_p $ & $ 1.43 ^{+ 0.31 }_{- 0.29 } $  $\rho_J$ \\

Planet temp ($A=0$)  & $ T_{\mbox{eql}} $ & $ 1119 ^{+ 26 }_{- 28 } $  K \\
& & \\
& & \\

$\chi^2_\nu$ (photometric) & $\chi^2_{phot}$    & 4145   \\
Photometric data points & N$_{phot}$ & 4151  \\
$\chi^2_\nu$ (spectroscopic) & $\chi^2_{spec}$   & 17.2  \\
Spectroscopic data points & N$_{spec}$ & 14  \\

\hline\\
\end{tabular}
\end{table}

\section{Discussion}

 Photometric surveys have now provided a large sample of transiting ESP
that can be used to determine their mass-radius relation and provide constraints
on their compositions.  
Here we presented the discovery of a new ESP with a mass of 2.96M$_J$, 
1.28R$_J$ radius, and a significant eccentricity of $0.059 ^{ + 0.014}_{ -0.004}$. 
We now discuss  the properties of WASP-10b in relation to the current sample
of transiting ESP, starting with its non-zero eccentricity.
 
%

Most of the current sample of published transiting ESP have orbits 
consistent with being circular and are fit with models using
zero eccentricity as is expected for short-period planets in orbits 
with semi-major
axes $<$ 0.2 AU.  Recent work \citep{jack08, mard07} 
has investigated the effects of tidal dissipation on the orbits of short period ESP.
The evolution of the orbital eccentricity appears to be driven primarily
by tidal dissipation within the planet, giving a circularisation  
timescale substantially less than 1 Gyr for typical tidal dissipation parameter,
Q$_p$ = 10$^5$ to 10$^6$. 
WASP-10 is a K dwarf with a spin period of 12 days and J$-$K=0.62 
and is rotating more slowly than stars of comparable colour in the 
Hyades \citep{tern00}. 
This suggests a rotational age between 600 Myr and 1 Gyr.
Thus, the persistence of
substantial orbital eccentricity in WASP-10b is therefore surprising. 

One plausible mechanism for maintaining the high eccentricity is secular  
interaction with an additional planet in the system. 
\citet{AL06} explore the effects of dynamical interactions among
planets in extrasolar planetary systems and conclude outer planets can
cause the inner planet to move through a range of eccentricities
over timescales that are short when compared to the lifetime of the
system, but very long when compared to the current observational baseline.
However, recently \citet{mat08} have argued that an unseen companion
driving short-Period systems is unlikely. They present an upper limit of
1 M$_{Neptune}$ for a possible unseen companion in the GJ 436 system and exclude
this based on the current radial velocity upper limits of $\leq$ 5 m/s. 
\citet{mat08} also  present a range of tidal quality Q$_p$ 
timescales that could be as large
as 10$^9$ years, and argue that this new class of eccentric, short period
ESP are simply still in the process of circularizing.  
WASP-10b has not been extensively studied to rule out a putative outer 
plane that may be driving its eccentricity. Thus,  
the $\approx$6\% eccentricity of WASP-10b makes it
an attractive target for future transit-timing variation studies, and  
for longer-term RV monitoring to establish the mass and period of the putative outer planet.


The majority of transiting ESP found have masses below 1.5M$_J$, 
although there are a few more massive ESP.
HD 17156, and COROT-Exo-2 have similar masses to WASP-10b and 
although there are two more massive ESP, the nearly 9 M$_J$ HAT P-2 (HD~149026b) \citep{bakos07} and
7.3 M$_J$ WASP-14b \citep{yog08}, this higher mass region has been poorly 
explored.
Additional transiting objects in the mass range are important for 
completing the current
ESP mass-radius relations and constraining their compositions. 
The current sample of transiting extrasolar giant planets (ESP) 
reveals a large range of densities.  We derive a mean density for WASP-10b 
of $\approx$1.89 g cm$^{-3}$ (1.42 $\rho_J$) and it would lie 
along the higher density contour in a mass-radius plot \citep{P08, sozz07}. 

One ultimate goal of our transit-search programme
is to provide the observational grist that will stimulate and advance refined models
for the formation and evolution of the hot and very hot Jupiters \citep{burr97, fortney2007, seag07}.  By
thus constraining the underlying physics, we will have a richer context for
the interpretation of the lower mass planets expected from missions such
as COROT and Kepler.

\section*{Acknowledgments}

The SuperWASP Consortium consists of astronomers primarily from the 
Queen's University Belfast, St Andrews, Keele, Leicester, The Open 
University, Isaac Newton Group La Palma and Instituto de  
Astrof{\'i}sica de Canarias. The SuperWASP Cameras were 
constructed and operated with funds made available from Consortium 
Universities and the UK's Science and Technology Facilities Council. 
SOPHIE observations have been funded by the Optical Infrared
Coordination Network. The data from the Mercator and NOT telescopes was 
obtained under the auspices of the International Time of the Canary 
Islands. We extend our thanks to the staff of the ING and OHP for their 
continued support of SuperWASP-N and SOPHIE instruments. FPK is
grateful to AWE Aldermaston for the award of a William Penney Fellowship.

\bsp



\label{lastpage}

\end{document}